\documentclass[%
twocolumn,
superscriptaddress,
longbibliography,
aps,
prl
]{revtex4-2}

\usepackage{amsmath,amssymb,bm,braket,url,dsfont,xspace,enumitem}
\usepackage{graphicx}
\usepackage{xcolor}
\usepackage{array}
\usepackage{booktabs}

\usepackage{longtable}

\newcolumntype{L}{>{$}l<{$}} 
\newcolumntype{C}{>{$}c<{$}} 

\usepackage[
draft=false,
bookmarks=true,
bookmarksnumbered=false,
bookmarksopen=false,
bookmarksopenlevel=4,
colorlinks=true,
anchorcolor=black,
citecolor=blue,
filecolor=magenta,
linkcolor=blue,
linkbordercolor={1 0 0},
urlcolor=blue,
pdfborder={0 0 1},
pdftitle={},
pdfauthor={},
pdfsubject={},
pdfkeywords={},
pdfpagemode=UseThumbs,
]{hyperref}

\usepackage{orcidlink}

\newcommand{\bk}{{\bm{k}}}

\newcommand{\T}{$\theta$\xspace}
\newcommand{\PT}{$\theta I$\xspace}
\newcommand{\Pa}{$I$\xspace}

\begin{document}

\title{
Ferroaxial Magnets: Time-reversal-even Mirror Symmetry Violation from Spin Order 
}

\author{Hikaru Watanabe
        \orcidlink{0000-0001-7329-9638}
        } 
\affiliation{Department of Applied Physics, Hokkaido University, Sapporo, Hokkaido 060-8628, Japan}

\author{Yue Yu
        \orcidlink{0000-0002-9446-9971}
        } 
\affiliation{Department of Physics, University of Wisconsin-Milwaukee, Milwaukee, Wisconsin 53201, USA}
\affiliation{School of Physics and Astronomy, University of Minnesota, Minneapolis, Minnesota 55455, USA}

\author{Jin Matsuda
\orcidlink{0000-0001-7329-9638}
        } 
\affiliation{Department of Applied Physics, The University of Tokyo, Hongo, Bunkyo-ku, Tokyo 113-8656, Japan}

\author{Daniel F. Agterberg
        \orcidlink{0000-0003-0178-1374}
        } 
\affiliation{Department of Physics, University of Wisconsin-Milwaukee, Milwaukee, Wisconsin 53201, USA}

\author{Ryotaro Arita
        \orcidlink{0000-0001-5725-072X}
        } 
\affiliation{Department of Physics, University of Tokyo, Tokyo 113-0033, Japan}
\affiliation{Center for Emergent Matter Science, RIKEN, {Wako}, Saitama 351-0198, Japan}

\begin{abstract}

        We investigate ferroaxial magnets, a new class of spin-order-driven multiferroic magnets in which magnetic ordering induces mirror-symmetry breaking while preserving both time-reversal and spatial-inversion symmetries.
        These systems exhibit a ferromagnet-like axial anisotropy that allows optical control of the ferroaxial polarization, while their macroscopic time-reversal symmetry makes them attractive for antiferromagnetic spintronics. Using spin crystallographic group analysis, we identify the candidate materials and the nonrelativistic ferroaxial nature stemming from the strong exchange splitting of magnets.
        Furthermore, a symmetry-based identification shows magnetic materials that host ferroaxial order and metallic conductivity, realizing the ferroaxial metal state that undergoes a ferroaxial phase transition while remaining metallic.
        As a direct probe for the ferroaxial metal, we propose a third-order nonlinear Hall effect originating from the transverse coupling between the electric field and Berry curvature dipole mediated by the ferroaxial anisotropy.
        Our results establish ferroaxial magnets as a platform for nonrelativistic multiferroicity and spintronic applications.

\end{abstract}

\maketitle


\noindent \textit{Introduction---}
Spintronic phenomena have been extensively explored in collinear and noncollinear magnetic materials.
Prominent properties include the responses of altermagnetic and noncollinear magnets~\cite{Jungwirth2025-symmetryreview,Naka2025-ai,Hayami2024-gd,Guan2022-yr,Smejkal2022-rk,Cheong2024-ak}.
The emergent spin-charge coupling arises from spin order without  relativistic spin-orbit coupling (SOC), indicating nonrelativistic realizations of spin-charge interconversion such as the spin-splitter and photomagnetic effects~\cite{Jungwirth2025-ms}.
Of particular interest is that the typical energy scale of the spin order is as large as the electrons' Coulomb interaction~\cite{Smejkal2022-ga,Guo2023-hf}.
Intriguing responses, which have previously been considered to stem from relativistic SOC, motivate a further material search covering weak spin-orbit-coupled cases such as  those based on $3d$ transition metal elements.  

Recent studies have further unveiled that spin order nonrelativistically offers physical consequences even without lifting spin degeneracy; \textit{e.g.}, anomalous Hall effect and nonreciprocal electric transport in spin-degenerate noncoplanar magnets~\cite{Martin2008-gk,Feng2020-qv,Watanabe2024-scg,Zhu2025-zd,Hayami2022-wt,Khanh2025-tk}.
For collinear antiferromagnets, time-reversal (\T{}) symmetric or parity-time-reversal (\PT{}) symmetric cases, which are termed $q$ magnets and \PT{} symmetric magnets~\cite{Watanabe2024-hw,Jin2025-bn8g}, do not show spin splitting and were classified as conventional antiferromagnets in stark contrast to altermagnets~\cite{Smejkal2022-ir}.
Despite the lack of spin-momentum locking, nontrivial macroscopic symmetry violations can offer emergent responses such as  natural optical activity and the nonlinear Hall effect through nonsymmorphic crystal symmetry and spin order with nonzero propagation vector $\bm{q}$~\cite{Jin2025-bn8g,Yu2025-k6b2}.

        \begin{figure}[htbp]
        \centering
        \includegraphics[width=0.8\linewidth,clip]{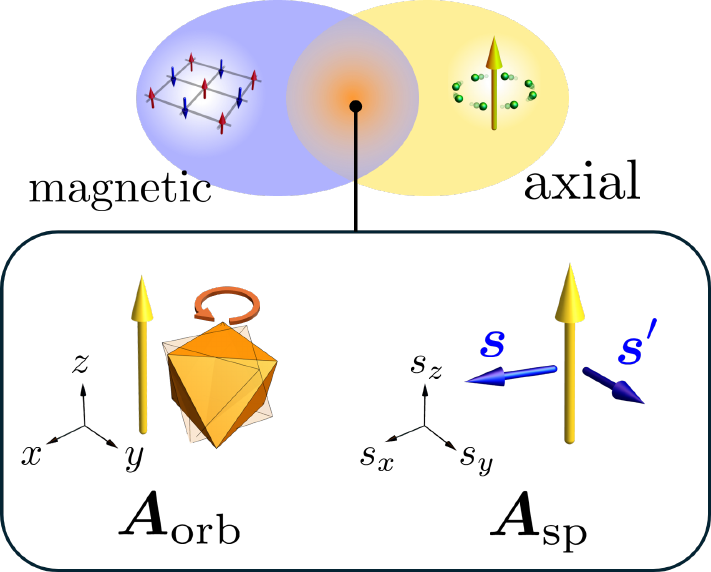}
        \caption{
                Sketch of axial-magnetic multiferroicity given by the combination of (antiferro-)magnetic and ferroaxial order (upper panel).
                In the absence of relativistic SOC, the multiferroic state can be classified by the orbital-active ($\bm{A}_\text{orb}$) and spin-active ($\bm{A}_\text{sp}$) ferroaxial anisotropy, which are defined by the \T{}-even axial vector in the real space ($x,y,z$) and spin space ($s_x,s_y,s_z$), respectively.
                $\bm{A}_\text{orb}$ can be active both in collinear and noncollinear magnets, while $\bm{A}_\text{sp}$ in noncollinear magnets.
                }
        \label{Fig_axialtype}
        \end{figure}

Prior studies have elaborated on nonrelativistic phenomena in magnetic materials with or without spin splitting.
They focus on magnetic materials exhibiting the \T{} or/and space-inversion (\Pa{}) symmetry breaking, whereas the \T{} and \Pa{} symmetric cases have not been largely explored.
Although spin splitting is forbidden, such \T{}-, \Pa{}-even symmetry violation is also of interest.
A promising example is  ferroaxial order, that is the ferroic order of \T{}-even axial vector $\bm{A}$~\cite{Naumov2004-bm,Prosandeev2006-vz}.
Ferroaxial order is realized in metamaterials and nonmagnetic insulators such as RbFe(MoO$_4$)$_2$,  NiTiO$_3$, and $1T$-TaS$_2$.
Its axial anisotropy means the lack of symmetry regarding mirror operations whose mirror plane is parallel to the ferroaxial vector $\bm{A}$.    
This mirror asymmetry has attracted much attention and has been investigated in optical measurements such as second-harmonic generation, electrogyration, and natural Raman optical activity.

While experimental investigations of ferroaxial physics have so far been carried out in nonmagnetic materials, the order can originate from spin order, as in the case of magnetoelectric multiferroics~\cite{Tokura2014-ix}. 
A seminal study investigated this new class of multiferroicity by magnetic point group analysis~\cite{Yang2025-ic}.
However, the nonrelativistic aspect of the spin-order-induced ferroaxiality has not been investigated.
From the perspective of SOC-free physics, a central question is whether ferroaxial anisotropy driven by spin order relies critically on SOC, which is a key issue for investigating the physical properties unique to this multiferroic state.

To this end, we present the spin crystallographic analysis of ferroaxial magnets.
The spin crystallographic group, the symmetry of magnets without relativistic SOC, allows us to identify the order parameter characterizing the ferroaxial magnet, that is, orbital- and spin-active ferroaxial vectors (Fig.~\ref{Fig_axialtype}).
We identify candidate materials hosting nonrelativistic multiferroicity and propose the possibility of ferroaxial materials with metallic conductivity, namely the \textit{ferroaxial metal}.
Furthermore, it is shown that the nonlinear transport governed by the Berry connection polarizability will be a fingerprint of the ferroaxial metal state.
Finally, we propose that the ferroaxial magnet serves as a promising material for antiferromagnetic and optospintronic applications due to its robustness against magnetic field as well as controllability by circular light.
The compatibility of two advantageous properties originates from its \T{}-even axial nature in sharp contrast to ferromagnets and anomalous Hall antiferromagnets~\cite{Smejkal2022-ga}.


\noindent \textit{Emergent ferroaxiality driven by spin order---}
Let us elaborate on the ferroaxial order invoked by spin order.
The ferroaxial motif can be given by the cross product of two vectors sharing the same space and time symmetries; \textit{e.g.}, $\nabla \times \bm{P}$~\cite{Jin2020-xk} where both the spatial gradient $\nabla$ and the electric polarization $\bm{P}$ are \T{}-even and polar vectors.
As in the case of magnetoelectric multiferroic magnets, the ferroaxial anisotropy can arise secondarily by the spin order.
We discuss candidate materials for ferroaxial magnets in the framework of the magnetic and spin space group analyses.

In the magnetic space group analysis, we classify the magnetic materials listed in \textsc{magndata}~\cite{Gallego2016-yz} by the ferroaxial vector $\bm{A}$.
To eliminate the trivial ferroaxial anisotropy in the crystal structure, we compare the rank of $\bm{A}$ ($\text{rank\,}\bm{A}$) in the paramagnetic and magnetic phases and identify the cases where $\text{rank\,}\bm{A}$ increases across the spin ordering.

The classification based on \textsc{spinspg} and \textsc{spglib}~\cite{Shinohara2023-qc,Togo2024-jw,spinspg} results in the identification of 19 ferroaxial magnets in total (see Ref.~\cite{Supplemental}).
Note that we focus on the candidate materials whose $n(\geq 3)$-fold rotation symmetry holds.
These candidates are well-suited for investigating the ferroaxial state, since the symmetry violation is not admixed with other anisotropies such as nematicity.
The identified materials include not only insulating oxides such as YMnO$_3$~\cite{Brown2006-pc}, but also metallic systems such as TmPdIn~\cite{Baran2016-om}, leading to the realization of ferroaxial metal.

We further investigate the symmetry of ferroaxial magnets using spin crystallographic symmetry~\cite{Watanabe2024-scg,Jungwirth2025-symmetryreview}.
The spin space group describes approximate symmetries for which the rotation operations separately act upon the spin and orbital space~\cite{Brinkman1966-qg,Opechowski1986}.
Then, the ferroaxial anisotropy can be attributed to either spin or orbital space.
The orbital ferroaxial vector $\bm{A}_\text{orb}$ is given by the orbital degrees of freedom like $\nabla \times \bm{P}$, while the spin part $\bm{A}_\text{sp}$ breaks the mirror symmetry in the spin space.
The latter is related to the even-parity vector spin chirality $\bm{A}_\text{sp} \sim \bm{s} \times \bm{s}'$ or the $p$-type spin nematic state~\cite{andreev1984spin,Chandra1991-cl,Yu2025-k6b2}, similarly to the spin octupolar order ($s_xs_ys_z$) arising from the all-in all-out spin order~\cite{Watanabe2024-hw}.
When relativistic SOC is taken into account, both $\bm{A}_\text{orb}$ and $\bm{A}_\text{sp}$ undergo the same transformation by a given operation, leading to spin-orbital-entangled ferroaxial order such as $\bm{A} \sim \bm{l} \times \bm{s}$ with the orbital magnetic moment $\bm{l}$~\cite{Hayami2022-yq}.
Owing to the noncollinear nature of $\bm{A}_\text{sp}$, the spin-active ferroaxial order arises in noncollinear magnets, while the orbital ferroaxial anisotropy $\bm{A}_\text{orb}$ can appear even in collinear magnets as in the case of (\T{}-even) odd-parity anisotropy~\cite{Perez-Mato2016-oj,Jin2025-bn8g,Yu2025-k6b2,Cao2024-arxiv}.

These two types of ferroaxial order give rise to characteristic physical phenomena without the help of SOC; \textit{i.e.}, orbital-active ferroaxial magnets show emergent responses related to orbital degrees of freedom such as natural Raman optical activity, while spin-active ferroaxial magnets host spin-related responses~\cite{YueYu-work}. 
Among the 19 candidate materials identified by the magnetic space group analysis, which is consistent with the prior study~\cite{Yang2025-ic}, the orbital active cases are CsCoF$_4$, FeTa$_2$O$_6$, CaFe$_3$Ti$_4$O$_{12}$, and the spin-active counterpart is CaCu$_3$Ti$_4$O$_{12}$~\cite{Gallego2016-yz}.
FeF$_3$, Na$_2$MnTeO$_6$, TmPdIn, and CaMn$_3$V$_4$O$_{12}$ are characterized by both orbital and spin active ferroaxiality.
We also perform the representation analysis for the orbital-active ferroaxial magnets and classify the candidate crystal and spin structures hosting $\bm{A}_\text{orb}$ in Ref.~\cite{Supplemental}, which will be convenient for further explorations of candidates.


\noindent \textit{Orbital-active ferroaxial physics---}
This work focuses on orbital-active ferroaxial order and associated physical phenomena.
Previous experimental studies are for optical phenomena of insulators.
 However, ferroaxial anisotropy has recently been reported to give rise to unconventional transport phenomena~\cite{Nasu2022-gu,Hayami2023-gv,Kurumaji2023-rb,Nakamura2024-of,Nishihaya2025-qv,Day-Roberts2025-ga}.
Known candidate materials undergoing ferroaxial order are limited to insulators, whereas the ferroaxial magnets we propose here offer a playground for  ferroaxial-order-induced transport phenomena.

Ferroaxial anisotropy relates physical quantities having the same space-time symmetry in the transverse manner as demonstrated in thermoelectric transport~\cite{Nasu2022-gu}, the unconventional Hall effect~\cite{Hayami2023-gv}, and nonlinear magnetic response~\cite{Inda2023-uw,Du2025-kz}.
This is in contrast to ferromagnetic order by which the transverse correlation occurs between the quantities with opposite \T{} parity; \textit{e.g.}, electric field and current for anomalous Hall effect.
In this work, making use of the ferroaxial transverse correlation, we demonstrate that the nonlinear Hall response is a hallmark of the ferroaxial metal state.

The (dc) nonlinear Hall response of our interest reads
                \begin{equation}
                J_a = \sigma_{a;bcd} E_b E_c E_d,
                \end{equation}
where electric current $\bm{J}$ is proportional to $|\bm{E}|^3$.
Nonlinear response has been applied to investigate the nontrivial \T{}-odd multipolar state and van der Waals materials~\cite{Lai2021-ht,Liu2022-fj,Wang2022-et,Li2024-xq,Fang2024-ed,Sorn2024-st}
Initially, we consider the highly symmetric ferroaxial symmetry denoted by the point group $\infty / m$.
The ferroaxial components are
                \begin{align}
                &\sigma_{x;yzz} = -\sigma_{y;xzz},\\
                &\sigma_{x;yyy} = - \sigma_{y;xxx} = \sigma_{x;xxy} = -\sigma_{y;xyy},
                \label{nonlinear_hall_planar}
                \end{align}
where we eliminated those allowed in the biaxial point group $\infty /mm$.
The former components can be considered as the $E_z^2$ correction to the Hall response in the $xy$ plane.
The latter components represent the nonlinear Hall effect confined to the basal plane and are dissipationless, as they are not totally symmetric with respect to the indices~\cite{Tsirkin2022-zf}.
We focus on this response because its dissipationless nature and transverse character within the basal plane make it particularly suitable for probing the ferroaxial metallic state.

The response originates from the field-induced Berry curvature dipole (BCD)~\cite{Sodemann2015,Matsyshyn2019-kd}.
BCD, that is, the odd-parity coupling between the Berry curvature and crystal momentum, is key to the second-order nonlinear Hall effect in the \T{} symmetric acentric systems.
Letting the momentum ($\bk$) resolved Berry curvature be $\bm{\Omega}^{(n)} (\bk)$ for the electronic band $n$, we define BCD as $D_{ab} = \int  \sum_n \partial_a \Omega_b^{(n)} (\bk) f_n \, d\bk /(2 \pi)^d $, in which $d$ is system's dimension and $f_n$ is Fermi-Dirac distribution function of energy $\varepsilon_n $.
The BCD-driven nonlinear Hall response is given by $J_a = \sigma_{a;bc} E_b E_c$ where
        \begin{equation}
                \sigma_{a;bc} = - \frac{q^2 \tau}{2 }    \epsilon_{abp} D_{cp} + \left[ b \leftrightarrow c \right],
                \label{BCD_nonlinear_hall}
        \end{equation}
where $q$ is the electron charge and $\tau$ is the relaxation time~\cite{deyo2009semiclassical,Moore2010,Sodemann2015}.

    \begin{figure}[htbp]
        \centering
        \includegraphics[width=0.9\linewidth,clip]{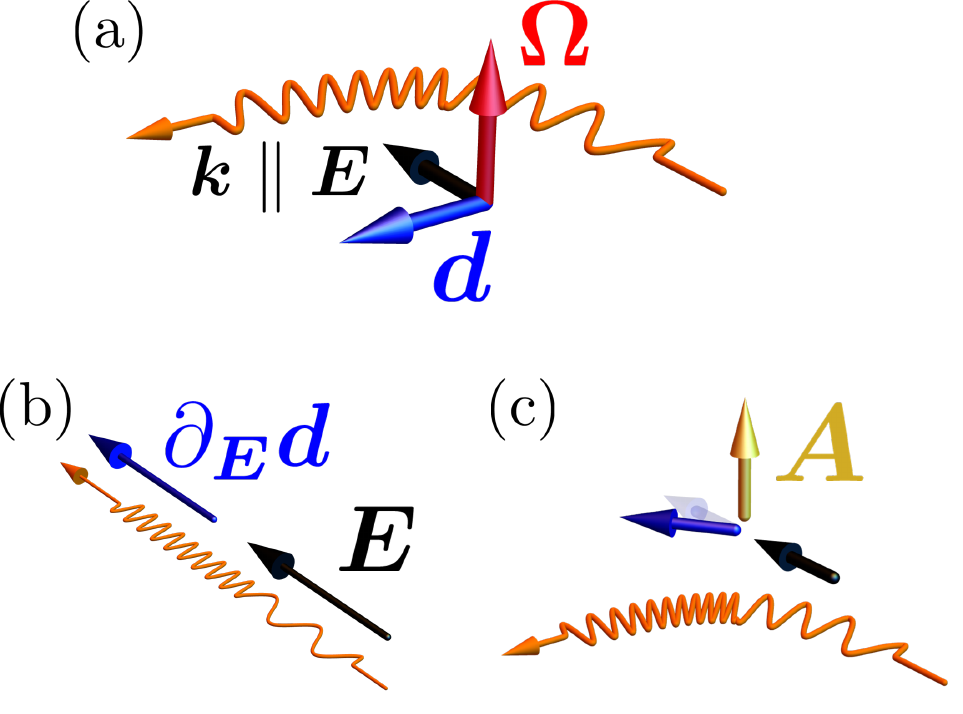}
        \caption{
                (a) Second-order nonlinear Hall effect originated from the Berry-curvature-dipole (BCD) vector $\bm{d}$ defined by the cross product of the momentum asymmetry ($\bk$) and Berry curvature $\bm{\Omega}$.
                The current (colored in orange) is rectified along $\bm{d}$, and thus $\bm{d} \perp \bm{E}$ is required for the Hall response.
                (b) Third-order nonlinear current response in an isotropic system.
                Field-induced BCD $\partial_{\bm{E}}\bm{d}$ is parallel to the electric field $\bm{E}$ [$d_0$ in Eq.~\eqref{ferroaxial_bcd}], leading to no Hall response.
                (c) Third-order nonlinear current response in the presence of ferroaxial order ($\bm{A}$).
                Owing to the lateral correlation between $\bm{E}$ and $\partial_{\bm{E}} \bm{d}$ by $\bm{A}$, the tilting component $d_\text{ax} \bm{A} \times \bm{E}$ allows for the Hall response as in the panel (a).
                }
        \label{Fig_hallcomparison}
        \end{figure}

The local Berry curvature vanishes when \PT{} symmetry is intact.
Accordingly, the ferroaxial system essentially hosts zero BCD, whereas the field-induced \Pa{} breaking steadily induces BCD as in the case of gate-induced Rashba spin-momentum locking.
The phenomenon is formulated by $\alpha_{ab}^{(n)} (\bk) = \partial \Omega^{(n)}_a (\bk) / \partial E_b = \epsilon_{apq} \, \partial ^2 \xi^{(nn)}_q (\bk) / \partial E_b \partial k_p $ with the Berry connection $\bm{\xi}^{(nm)}$ for the band indices $(n,m)$.
The derivative $\partial \xi^{(nn)}_{q}/ \partial E_b$ is called Berry connection polarizability (BCP)~\cite{Gao2014}.
BCP leads to the field-induced BCD as
                \begin{equation}
                \frac{ \partial D_{ab}}{\partial E_c} =   \int \frac{d\bk}{ (2\pi)^d} \sum_n  \frac{\partial \alpha_{b c}^{(n)}(\bk)}{\partial k_a} f_n.
                \end{equation}
Plugging the induced BCD into Eq.~\eqref{BCD_nonlinear_hall}, we obtain the third-order nonlinear Hall response as
                \begin{equation}
                \sigma_{a;bcd} = - \frac{q^3 \tau}{2 }    \epsilon_{abp} \frac{\partial D_{cp}}{\partial E_d} + \left[ b \leftrightarrow c \right].
                \label{third_order_nonlinear_hall}
                \end{equation}
BCP is explicitly given by~\cite{Gao2014}
                \begin{equation}
                        G_{ab}^{(n)} \equiv \frac{1}{q}\frac{\partial \xi_a^{(nn)}}{\partial E_b}   = 2 \text{Re} \sum_{m \neq n} \frac{\xi^{(nm)}_a \xi^{(nm)}_b}{\varepsilon_n - \varepsilon_m}.
                        \label{Berry_connection_polarizability}
                \end{equation}
Finally, we rewrite Eq.~\eqref{third_order_nonlinear_hall} as
                \begin{align}
                \sigma_{a;bcd} = - \frac{q^3 \tau }{3}
                &\left[  \Lambda_{ab;cd} + \Lambda_{ac;bd} + \Lambda_{ad;bc} \right. \notag \\
                & \left. - \Lambda_{bc;ad}  - \Lambda_{cd;ab}  - \Lambda_{bd;ac} \right].
                \label{final_formula_nonlinear_hall}
                \end{align}
Note the formula is symmetrized with respect to fields' indices ($b,c,d$)~\cite{Lai2021-ht,Liu2022-fj,Xiang2023-bv,Mandal2024-me}.
Owing to the key role of the quadrupole moment of BCP $\Lambda_{ab;cd} = \int [d\bk] \sum {}_n f_n \partial_a \partial_b G_{cd}^{(n)}$
, the contribution is called the BCP quadrupole mechanism~\footnote{
        Note that the induced current is not always parallel to the electric field for Eq.~\eqref{final_formula_nonlinear_hall}~\cite{Mandal2024-me}, though the transverse response is obtained in the ferroaxial case.
        We also note that the third-order nonlinear Hall effect can be obtained as the current-induced Berry curvature polarizability; \textit{i.e.,} for the electric-field-induced uniform Berry curvature $\delta \Omega_a = \chi_{ab} E_b$, the \Pa{}-odd and \T{}-odd response function $\hat{\chi}$ can be driven by the electric current as $\partial \chi_{ab}/ \partial J_c$.
        As in the case of second-order nonlinear responses~\cite{Gao2014,Watanabe2020-oe}, the uniform Berry curvature induced by both of current and field leads to the transverse transport.
        Owing to the Ohmic current, the response totally depends on $\tau^1$, similarly to Eq.~\eqref{final_formula_nonlinear_hall}.
        This contribution is derived with the full-quantum approach, but the dissipationless componnent differs from Eq.~\eqref{final_formula_nonlinear_hall} only by a scalar factor.
}.
We note that another contribution such as the Drude effect are excluded because of its dissipative nature.
Furthermore, the conserved \PT{} symmetry strongly suppresses extrinsic contributions such as the skew scattering effect~\cite{Watanabe2020-oe,Watanabe2024-hw}.

We demonstrate that ferroaxial anisotropy leads to the BCP-driven nonlinear Hall response.
First, we consider a highly symmetric case such as that labeled by $\infty / m m$.
As in the case of gate-controlled Rashba SOC, an applied electric field can induce BCD; \textit{i.e.}, the vector $d_a = -\epsilon_{abc} D_{bc}$  is induced along $\bm{E}$ as $\bm{d} \parallel \bm{E}$.
However, this induced BCD vector $\bm{d}$ is not desirable because the Hall response requires $\bm{d}\perp \bm{E}$ [Fig.~\ref{Fig_hallcomparison}(a,b)].

A striking feature of the ferroaxial system is that the BCD vector is inflected as
        \begin{equation}
        \bm{d} = d_0 \bm{E} + d_\text{ax} \bm{A} \times \bm{E},
        \label{ferroaxial_bcd}
        \end{equation}
since the (orbital) ferroaxial anisotropy leads to the transverse correlation between $\bm{E}$ and $\bm{d}$, both of which are \Pa{} odd and \T{} even.
For instance, under ferroaxial symmetry $\infty / m$ allowing $\bm{A} \parallel \hat{z}$, the field $\bm{E} \parallel \hat{x}$ induces the BCD vector $d_\text{ax} \hat{y}$ in addition to the trivial component $d_0 \hat{x}$, leading to the nonlinear Hall response $\sigma_{y;xxx}$ [Fig.~\ref{Fig_hallcomparison}(c)].
As is evident from Eq.~\eqref{ferroaxial_bcd}, the obtained Hall coefficient changes its sign depending on the ferroaxial polarization.
Furthermore, the response is allowed in the presence of the Fermi surface and hence is characteristic of ferroaxial metals.


        \begin{figure*}[htbp]
        \centering
        \includegraphics[width=0.8\linewidth,clip]{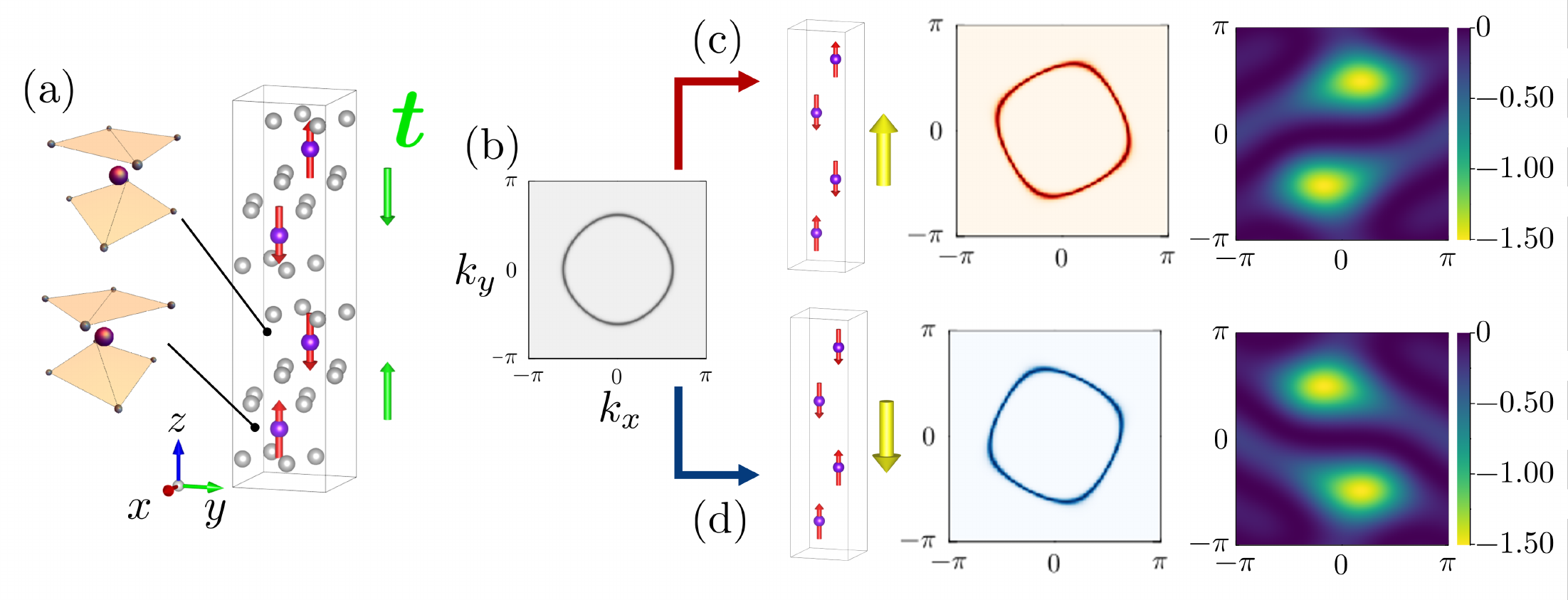}
        \caption{
                (a) Crystal and spin structures for the model study (right panel).
                The purple balls are magnetic sites surrounded by gray-colored ligands, which define the staggered chirality of magnetic sites (left panel).
                The spin order doubling the unit cell to the $z$ axis allows for the toroidal moment $\bm{t}$ in each original unit cell due to the coupling of the spin magnetic moment and chirality.
                (b) Fermi surface ($k_z=0,~\mu=-1.2$) without the nonsymmorphic symmetry [\textit{i.e.} local chirality depicted in (a)].
                (c,d) Spin-driven ferroaxial states illustrated by yellow-colored ferroaxial vectors (left panel), fermi surface ($k_z=0,~\mu=-1.2$) tilted from that in (b) due to the ferroaxial symmetry (center panel), and the Berry connection polarizability $G_{xx} (\bk)$ summed over the bands below $\mu=0$ (right panel).   
                }
        \label{Fig_model}
        \end{figure*}
\noindent \textit{Model study---}
Next, we microscopically demonstrate the ferroaxial Hall response.
The model is for a ferroaxial $q$ magnet in which the collinear spin order changes the point group symmetry from $4/mmm1'$ to $4/m1'$, thereby invoking $\bm{A} \parallel \hat{z}$.
The crystal structure is nonsymmorphic and tetragonal as $P4 / nnc$ (No.126), and the spin order is characterized by the propagation vector $\bm{q} = [0,0,\pi]$, as schematically illustrated in Fig.~\ref{Fig_model}(a). 
In stark contrast to typical conventional collinear antiferromagnets~\cite{Smejkal2022-ga}, the nontrivial coupling between the nonsymmorphic operation and spin order with $\bm{q} \neq \bm{0}$ gives rise to macroscopic symmetry breaking, similarly to improper magnetoelectric multiferroicity~\cite{Perez-Mato2016-oj}.
The magnetic atoms are in a locally noncentrosymmetric environment~\cite{Sigrist2014-gu}; more specifically, they show chirality determined by surrounding ligands [Fig.~\ref{Fig_model}(a)]. 
The antiferroic spin order coupled with local noncentrosymmetry leads to the magnetic toroidal moment $\bm{t}$ inside the original unit cell~\cite{Watanabe2018-do,Hayami2018-bh}, and hence it is indicated that the alternating \PT{}-symmetric magnetic order leads to the ferroaxial order through the nonsymmorphic property [Fig.~\ref{Fig_model}(a)].

        \begin{figure}[htbp]
        \centering
        \includegraphics[width=0.9\linewidth,clip]{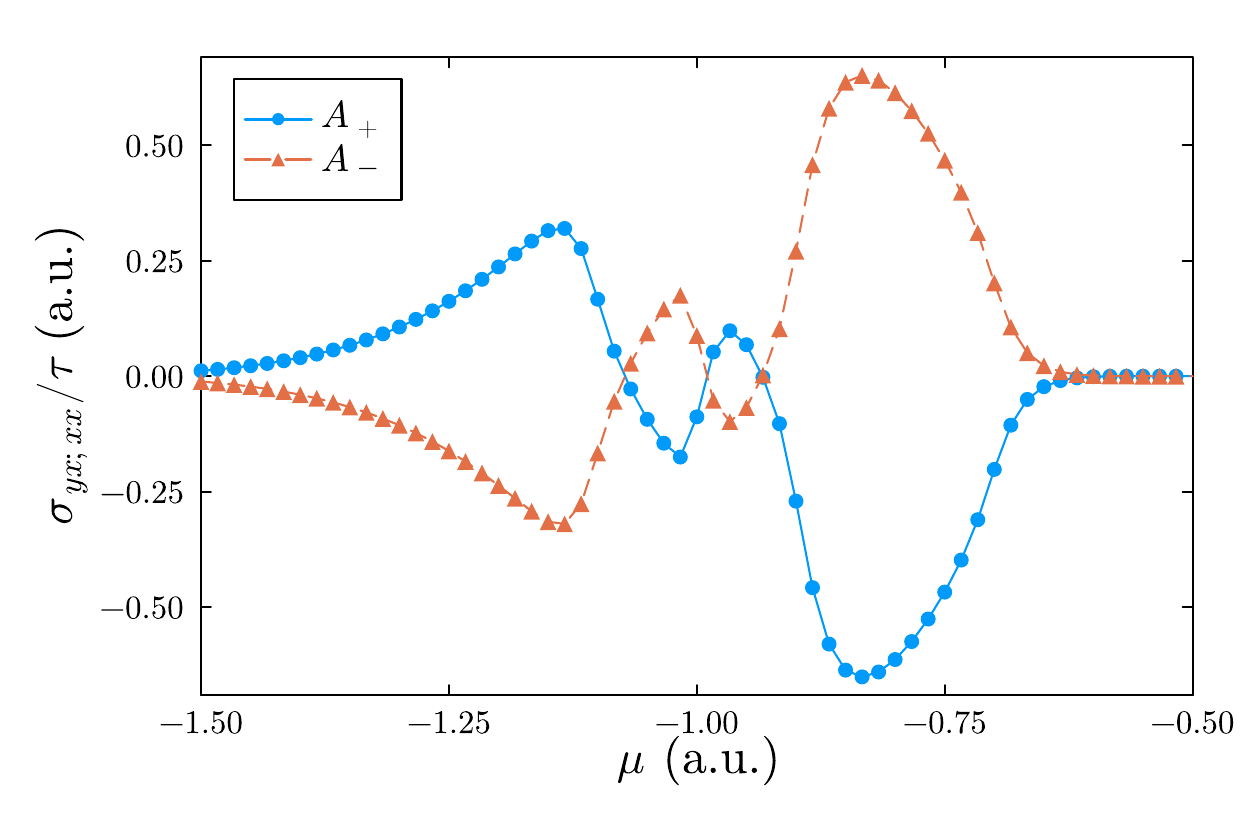}
        \caption{
                Third-order nonlinear conductivity $\sigma_{y;xxx}$ and its chemical potential ($\mu$) depedence.
                Two curves are for the spin-order-driven ferroaxial states with opposite polarizations ($A_\pm$).
                Note that a negligible response is obtained in the insulator regime ($\mu \gtrsim -0.6$).
                We used $q=1$, temperature $T=10^{-2}$, and the Brillouin zone meshing $N=100^3$ for the summation over $\bk$.
                }
        \label{Fig_NHE}
        \end{figure}

We employ the single orbital tight-binding model introduced in Ref.~\cite{Supplemental}.
The key ingredients are the hopping terms reflecting the nonsymmorphic structure and the exchange splitting by the collinear and $\bm{q} \neq \bm{0}$ spin order.
It is noteworthy that the relativistic SOC is not included and that the ferroaxial anisotropy can be significant even with weak SOC.
Furthermore, because the collinear spin state forbids the spin-active ferroaxial vector $\bm{A}_\text{sp}$, the model is appropriate for exploring the orbital ferroaxial order $\bm{A}_\text{orb}$.

The ferroaxial vector $\bm{A}$ is opposite between the spin-ordered states related by the out-of-plane mirror operation such as $m_{yz}$ [Fig.\ref{Fig_model}(c,d)].
The opposite ferroaxial polarizations are consistent with the Fermi surfaces of each state.
The swirling Fermi surface respects the four-fold rotation symmetry and spin degeneracy, while it does not have any of the out-of-plane mirror symmetries that exist in the paramagnetic state [Fig.~\ref{Fig_model}(b)].
The rotating behavior is similarly observed in BCP of the Eq.~\eqref{Berry_connection_polarizability}.
Specifically, as shown in Fig.~\ref{Fig_model}(c,d), the twisted distribution of $G_{xx}$ indicates the quadrupole moment $\Lambda_{xy;xx} \neq 0$, allowing for the Hall effect $\sigma_{yxxx}$.

As in the azimuthal-symmetric case,  ferroaxial order realizes the nonlinear Hall effect given by Eq.~\eqref{nonlinear_hall_planar}.
For $\sigma_{y;xxx}$ as an example, the formula reads
\begin{equation}
        \sigma_{y;xxx} = - \frac{q^3 \tau }{2} \left( \Lambda_{xy;xx} - \Lambda_{xx;xy} \right).
\end{equation}
We plot the chemical potential ($\mu$) dependence of $\sigma_{y;xxx}$ [Fig.~\ref{Fig_NHE}] for opposite ferroaxial polarizations $A_+ = -A_-$.
In agreement with the formula, the response is allowed in the metal regime, and the opposite signs are obtained as $\sigma_{y;xxx} (A_+) = -\sigma_{y;xxx} (A_-)$.


\noindent \textit{Discussions---}
This work elaborates the nonrelativistic realization of ferroaxial magnets and the possibility of ferroaxial metals.
Candidates include those hosting both metallic conductivity and ferroaxial order such as TmPdIn, which is in contrast to previously considered metals whose ferroaxial anisotropy is not due to spontaneous order but to crystal structure~\cite{Wimmer2015-xw,Yang2025-up}.
The identified candidates include conventional antiferromagnets, which have been considered less versatile, but our work further broadened the scope of functional antiferromagnets.
It is noteworthy that ferroaxial magnets are switchable by the circular-light drive, as in the case of the nonmagnetic ferroaxial materials~\cite{He2024-ie,Zeng2025-jr}.
By considering their magnetic-field tolerance ensured by the global \T{} symmetry as well, ferroaxial magnets may be promising for spintronic applications. 
Furthermore, the nonrelativistic nature implies that emergent properties characteristic of ferroaxial anisotropy can be enhanced by the strong Coulomb interaction of magnetic atoms as demonstrated in studies of spin-charge-coupled responses in \T{}-broken collinear and noncollinear magnets~\cite{Jungwirth2025-ms,Smejkal2022-rk}.

Lastly, the \T{}-symmetric macroscopic symmetry violation, such as ferroelecticity, chirality~\cite{Yu2025-k6b2,Jin2025-bn8g}, and ferroaxiality, can be found in the context of multipolar anisotropy.
In particular, our analysis further implies the magnetic realization of the \T{}-even axial octupolar order labeled by the Laue class $m\bar{3}1'$~\cite{Watanabe2025-bf,Supplemental}.
According to our classification, several candidates are found in \textsc{magndata}~\cite{Chamard-Bois1972-aj,Hirschmann2022-kf}.
The emergent response and controllability of multipolar order have recently been clarified in the context of nonmagnetic materials~\cite{Watanabe2025dualcircularramanopticalactivity}.
Moreover, following the parallel discussion to the ferroaxial case, we obtain the third-order current response; the response is determined by $\Lambda_{xx;yy}-\Lambda_{yy;xx}$ and its cyclic permutations, which is sensitive to the axial octupolar order. 
The exploration of such \textit{multipolar multiferroics} will also be promising. 
In summary, we have proposed the spin-order-driven ferroaxial order and the possibility of ferroaxial metal state, which can be sensed by the nonlinear Hall effect relating to the quantum geometry in ferroaxial states.

\section*{acknowledgement}
We thank Tsuyoshi Kimura for fruitful discussions.
H.W. thanks Fernando de Juan for his helpful comments on a ferroaxial material and thanks Katsuhiro Tanaka for his fruitful comments.
This work is supported by Grant-in-Aid for Scientific Research from JSPS KAKENHI Grant
No.~JP23K13058 (H.W.),
No.~JP24K00581 (H.W.),
No.~JP25H02115 (H.W.),
No.~JP21H04990 (R.A.),
No.~JP25H01246 (R.A.),
No.~JP25H01252 (R.A.),
JST-CREST No.~JPMJCR23O4(R.A.),
JST-ASPIRE No.~JPMJAP2317 (R.A.),
JST-Mirai No.~JPMJMI20A1 (R.A.),
and RIKEN TRIP initiative (RIKEN Quantum, Advanced General Intelligence for Science Program, Many-body Electron Systems).
D.F.A. and Y.Y. were supported by the Department of Energy, Office of Basic Energy Science, Division of Materials Sciences and Engineering under Award No DE-SC0021971.
H.W. was also supported by JSR Corporation via JSR-UTokyo Collaboration Hub, CURIE.
We used \textsc{spinspg}~\cite{spinspg} on top of \textsc{spglib}~\cite{Togo2024-jw,Shinohara2023-qc} for symmetry analyses and \textsc{vesta} for the visualization of crystal and spin structures~\cite{Momma2011-jl}.

\end{document}